\documentclass[preprint, twocolumn, 3p, sort&compress, longtitle]{elsarticle}
\usepackage{epsfig}
\usepackage{amsmath}
\usepackage{amsfonts}
\usepackage{amssymb}
\usepackage{graphicx}
\usepackage{colordvi}
\usepackage{color}

\journal{Physic Letters B}
\begin{document}
\begin{frontmatter}
\title{On the universality of MOG weak field approximation at galaxy cluster scale}
\author[idm]{Ivan De Martino\corref{cor1}}
\ead{ivan.demartino@ehu.eus}
\author[mdl1,mdl2,mdl3]{Mariafelicia De Laurentis}
\ead{laurentis@th.physik.uni-frankfurt.de}
\cortext[cor1]{Corresponding author}
\address[idm]{Department of Theoretical Physics and History of Science, 
University of the Basque Country UPV/EHU, Faculty of Science and Technology, 
Barrio Sarriena s/n, 48940 Leioa, Spain}

\address[mdl1]{Institute for Theoretical Physics, Goethe University, Max-von-Laue-Str.~1, D-60438 Frankfurt, Germany}
\address[mdl2]{Tomsk State Pedagogical University, ul. Kievskaya, 60, 634061 Tomsk, Russia}
\address[mdl3]{Lab.Theor.Cosmology,Tomsk State University of Control Systems and Radioelectronics (TUSUR), 634050 Tomsk, Russia}


\begin{abstract}
In its weak field limit, Scalar-tensor-vector gravity theory introduces
a Yukawa-correction to the gravitational potential. Such a correction depends on the two parameters, 
$\alpha$ which accounts for the modification of the gravitational constant, and $\mu^{*-1}$ which
represents the scale length on which the scalar field propagates. These parameters were found to 
be universal when the modified gravitational potential was used to fit the galaxy rotation curves and the mass
profiles of galaxy clusters, both without Dark Matter. We test the universality of these parameters 
using the the temperature anisotropies due to the thermal Sunyaev-Zeldovich effect. In our model 
the intra-cluster gas is in hydrostatic equilibrium within the modified gravitational potential well 
and it is described by a polytropic equation of state. We predict the thermal Sunyaev-Zeldovich
temperature anisotropies produced by Coma cluster, and we compare them with those obtained 
using the Planck 2013 Nominal maps. In our analysis, we find $\alpha$ and the scale length, respectively, 
to be consistent and to depart from their universal values.
 Our analysis points out that the assumption of the universality 
of the Yukawa-correction to the gravitational potential  is ruled out at more than $3.5\sigma$ at galaxy clusters scale,
while demonstrating that such a theory of gravity is capable to fit the cluster profile if the scale dependence
of the gravitational potential is restored.
\end{abstract}
\end{frontmatter}

\section{Introduction}\label{I}

Scalar-Tensor-Vector Gravity theory (STVG), also known as MOdified Gravity (MOG), 
adds scalar, tensor and massive vector fields to the standard Hilbert-Einstein action  \cite{Moffat2006a, Moffat2006b}. 
In particular, the mass of the MOG vector field and its strength are governed by two running constants, 
$\alpha$ and $\mu^*$, that are promoted to scalar fields and can be constrained by data.

Similarly to $f(R)$ gravity \cite{PhysRept,Annalen,OdintsovPR},  MOG theory introduces a Yukawa-like correction
to the Newtonian gravitational potential in its weak field limit \cite{Moffat2006b}.
Specifically, the modified gravitational potential  is \cite{MoffatRahvar2013}:
\begin{align}
\Phi_{\rm eff}(\vec x) = - G_N\int\frac{\rho(\vec x')}{|\vec x-\vec x'|}\left[1+\alpha-\alpha e^{-{\mu^*}|\vec x-\vec x'|}\right]d^3\vec{x}',
\label{eq:MOG1}
\end{align}
where $\mu^*$ is the inverse of the characteristic length of the modified gravitational potential, 
that acts at a certain scale for the self-gravitating systems,
and $\alpha = (G_\infty - G_N)/G_N$ accounts the modification of
the Newton constant \cite{Moffat2007e}, where $G_\infty$ is the effective gravitational constant
at infinity.

At cosmological scales, MOG correctly predicts accelerated expansion of the universe 
and the emergence of the Large Scale structure \cite{Moffat2005, Moffat2007a, Moffat2007b, Moffat2009}. 
At much smaller scales, it is able to correctly predict the Tully-Fisher relation and 
the galaxy rotation curves \cite{Brownstein2006a,Brownstein2006} with 
$\alpha$ and $\mu^*$ being "universal" parameters with constrained values 
$\alpha=8.89\pm0.34$ and $\mu^*=0.042\pm0.004\,{\rm kpc}^{-1}$, 
respectively \cite{MoffatRahvar2013}. Despite its successes at galactic scale, 
it is not clear if the assumption of the universality of those parameters holds at the scale of galaxy clusters. 
In general these parameters depend on the mass of the source of the gravitational potential and, therefore, they should 
depend on the scale length of the self-gravitating system as their analogue in $f(R)$ gravity \cite{Annalen}. 
Nevertheless, the universal parameters seem to be able to predict the dynamical mass of galaxy clusters 
\cite{Brownstein2007, arXiv:1606.09128}.

In this letter we propose an alternative test to probe the universality of the $\alpha$ and $\mu^*$ parameters
at galaxy cluster scale.  We use {\it Planck} 2013 Nominal maps to measure 
the thermal Sunyaev-Zeldovich (TSZ) profile  \cite{tsz} and to constrain 
the parameters $(\alpha, \mu^*)$ of the modified gravitational potential. We focus
our analysis on the Coma cluster since it is located close to the galactic  pole where the
foreground emission is comparatively low. The letter is organized as follows: in Sec. \ref{II} we briefly
describe the data; in Sec. \ref{III} we illustrate the methodology used to fit the profile to the data;
in Sec. \ref{IV} we discuss the results of our analysis; in Sec. V we point out the limitation of our analysis
and the future perspectives in this field; and finally, in Sec. \ref{VI} we give our main conclusions.

\section{Data}\label{II}

In 2015 the Planck Collaboration made publicly available the Compton Y-maps \cite{planck15_22} that
was obtained applying a component separation algorithm to the high frequency channels (100-857 GHz) 
of the Planck mission. This technique extracts a signal when its frequency dependence is specified. 
Let us note that in order to specify the TSZ frequency dependence, one can not include any relativistic effect 
(due to the electron temperature) to the frequency dependence, and  have to fix the CMB temperature-redshift 
relation to be (adiabatic): $T_{CMB}(z)=T_{CMB}(0)(1+z)$. 
However,  the intra-cluster medium of Coma cluster has a temperature $T_e\sim7$ keV \cite{Vikhlinin2006} and 
relativistic effect contributes $\sim10\%$ to the total TSZ emission. 
Moreover, it is well known that alternative theories of gravity  could produce a departure from the 
 adiabatic expansion since they could change the evolution of  cosmological background and 
its density perturbations.

\begin{figure}
\epsfxsize=1.0\columnwidth \epsfbox{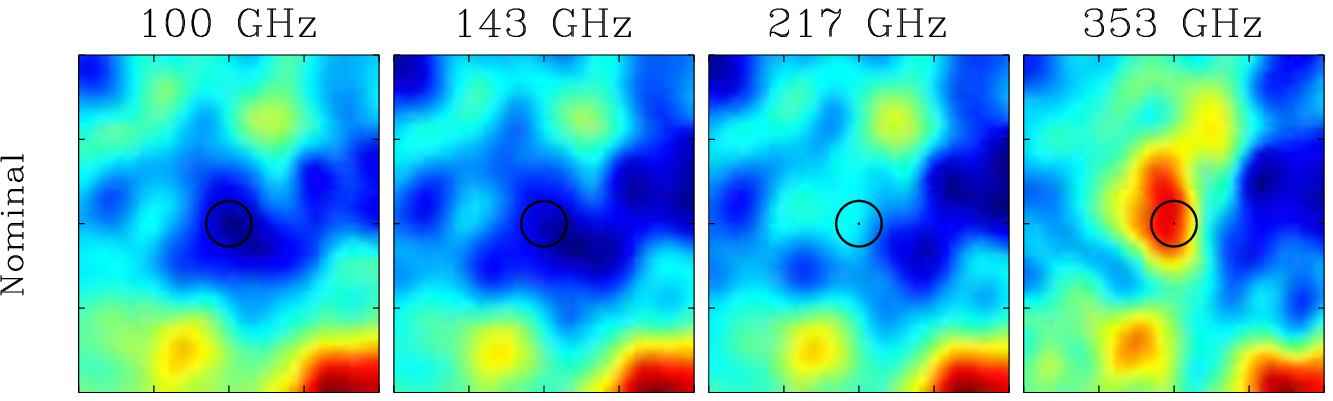}\\
\epsfxsize=1.0\columnwidth \epsfbox{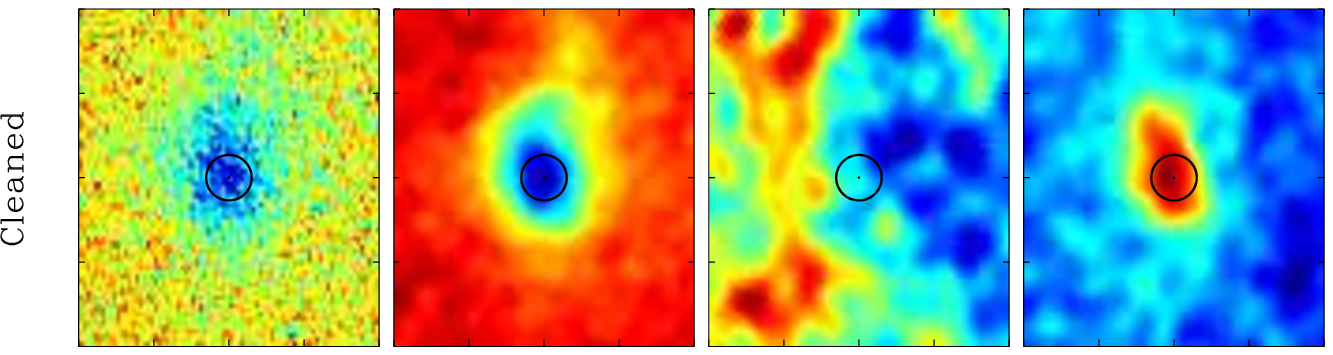}\\
\epsfxsize=1.0\columnwidth \epsfbox{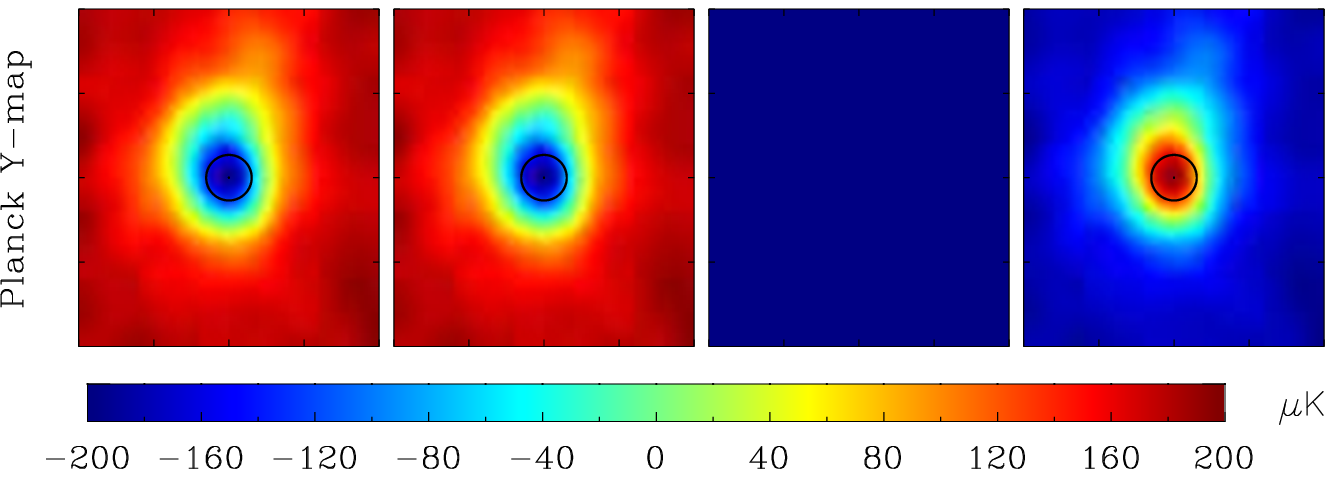}
\caption{Patches centered at the position of A1656 (Coma cluster) at 100-353 GHz. Patches are $4^\circ\times4^\circ$.
First, second, and third rows illustrate the view of Coma cluster in Planck Nominal, foreground cleaned, 
and Planck Compton maps, respectively. }\label{fig1}
\end{figure}

Therefore, although the Planck Y-maps allows to measure the SZ cluster profiles with few percent accuracy within its virial radius,
due to a lack of information about the effect of MOG theory at cosmological scales, 
we prefer to be more conservative and  test the underlying theory of gravity  by measuring the TSZ profile on 
the {\it Planck} Nominal maps \cite{planck13_I, gorski2005}.
However,  to reliably detect the TSZ temperature anisotropies induced by a galaxy cluster we need to reduce the
contaminations due to foreground emissions such as galactic dust, CO lines, synchrotron radiation, point and 
extended infrared sources, and the cosmological CMB signal. For that purpose, we applied the cleaning procedure 
described in \cite{demartino2015b, demartino2016a} to the high frequency channels. Briefly, the main steps are the following:
{\em (i)} maps were brought to a common 10 arcminutes resolution corresponding to the
angular resolution of the 100 GHz channel; {\em (ii)} CO lines were removed using the CO maps released by the Planck 
Collaboration \cite{planck13_13}; {\em (iii)} intrinsic CMB signal and kSZ were removed using an LGMCA template \cite{LGMCA1, LGMCA2};
{\em (iv)} the dust emission were removed by using the highest frequency channel as a template.
Finally, we measure the TSZ temperature anisotropies produced by Coma cluster at $100$, $143$, and $353$ GHz 
channels while we discard the $217$ and $545$ GHz channels: the first channel does not provide useful information since the TSZ signal
is greatly reduced ($\sim 0$); in the second one, residuals of the thermal dust emission are 
still the dominant contribution.  To show the effectiveness of our cleaning procedure we show in Figure \ref{fig1}
a $4^\circ\times4^\circ$ patch centered at the position of Coma cluster for the 100, 143, 217, 353 GHz channels. 
The first row shows the view of Coma in the Planck Nominal maps; the second row shows the results of our 
cleaning procedure and the last row shows the view of the galaxy cluster
in the Y-map released by the Planck Collaboration. The latter, to be compared to our cleaned data, has been
multiplied by the frequency dependence of the TSZ effect. Our cleaning procedure  produces a noisier map with more residuals. 
This is reflected in our error bars that are larger than the one obtained from the Planck Collaboration especially at
larger radii. However, we prefer to be more conservative and use our own data to test MOG theory for the reasons explained above.

To compute the error bars, we carried out  $1,000$ random simulations. In each one we placed a synthetic (Coma-like) cluster in a
random position in the sky, and we apply the cleaning procedure described above. Then, we measured the profile 
at the same angular apertures of the real cluster. The positions were taken out of the known galaxy clusters listed in 
in X-ray catalog \cite{Piffaretti2011}. Finally, we used the simulated profiles to compute the correlation matrix ($C_{ij}$) 
between different apertures, and we used the latter to compute the chi-square in our statistical analysis.

\section{Methodology}\label{III}

The TSZ temperature anisotropies are produced when CMB photons are scattered off
by the high energy electrons in the Intra-Cluster Medium.  Such anisotropies are
usually expressed as 
\begin{align}
\frac{\Delta T_{TSZ}(\hat{n})}{T_0}= G(\tilde{\nu})\frac{\sigma_{\rm T}}{mc^2}\int_l P_{\rm e}(l)dl.
\label{eq:y_c}
\end{align}
where $P_e(l)$  is pressure profile along the line of sight $l$, 
$T_0=2.725\pm 0.002$K is the present value of the CMB black-body 
temperature \cite{fixsen}, and $G(\tilde{\nu})$ is the spectral frequency dependence where
$\tilde{\nu}=h\nu(z)/k_BT(z)$ is the reduced frequency.
In the non relativistic limit (electron temperature about few keV), 
$G(\tilde{\nu})= \tilde{\nu}{\rm \coth}(\tilde{\nu}/2)-4$.  Relativistic corrections 
in the electron temperature up to fourth order have been included to improve the  
model \cite{itoh1998, nozawa1998, nozawa2006}. 

To predict the TSZ temperature anisotropies, the pressure profile must be 
specified. Following  \cite{demartino2014, demartino2016}, we considered 
the gas in hydrostatic equilibrium within the modified potential well 
of the galaxy cluster 
\begin{equation}\label{eq:HE}
\frac{d\mathbb{P}(r)}{dr}=-\rho_{gas}(r)\frac{d\Phi_{\rm eff}(x)}{dr},
\end{equation}
and well described by a polytropic equation of state 
\begin{equation}\label{eq:PES}
\mathbb{P}(r)\propto\rho_{gas}^\gamma(r).
\end{equation}
The system of equations is closed with the conservation of the mass
\begin{equation}\label{eq:EMC}
\frac{dM(r)}{dr}= 4\pi\rho_{gas}(r).
\end{equation}
Let us remark that the model does not include any Dark Matter component. 
Thus, the pressure profile  $P_e(r)=P_c\mathbb{P}(r)$ depends by the two MOG parameters  $(\alpha, \mu^*)$
the polytropic index $\gamma$,  and the central pressure $P_c$.
Finally, to predict the TSZ temperature anisotropies, 
the profile was integrated along the line of sight and 
convolved with the $10$ arcminutes antenna beam of the {\it Planck} data.

To test the universality of the MOG weak field approximation 
we predicted the TSZ profile from  $5$ to $100$ arcminutes (in rings of $5$ 
arcminutes width), and we fit them to the data at the same apertures
carrying out a Monte Carlo Markov Chain (MCMC) analysis employing 
the Metropolis-Hastings sampling algorithm and the Gelman-Rubin convergence 
criteria \cite{Metropolis1953, Hastings1970, Gelman1992}. 
We run four independent chains, each one composed by 25,000 steps, 
with randomly set starting points. 
The parameter space explored by our pipeline is given in Table \ref{tab:priors}.
\begin{table}
\begin{center}
\begin{tabular}{|ccc|}
\hline
 {\bf Parameter} & {\bf Priors} & {\bf References}\\
 \hline
 $P_c /[10^{-2}$ cm$^{-3}$ keV]  & $[0.0, 3.0]$ & \cite{demartino2016}\\
 $\gamma$ & $[1.0, 5/3]$  & \cite{demartino2016}\\ 
 $\mu^{*^{-1}}$/[Mpc] & $[0.01, 20.0]$  & \cite{MoffatRahvar2013, MoffatRahvar2014} \\
 $\alpha$ & $[0.1, 20.0]$  & \cite{MoffatRahvar2013, MoffatRahvar2014} \\
 \hline
\end{tabular}
\caption{Parameter space explored by the MCMC}\label{tab:priors}
\end{center}
\end{table}

\section{Results and Discussion.}\label{IV}

\begin{figure}
\epsfxsize=1.0\columnwidth \epsfbox{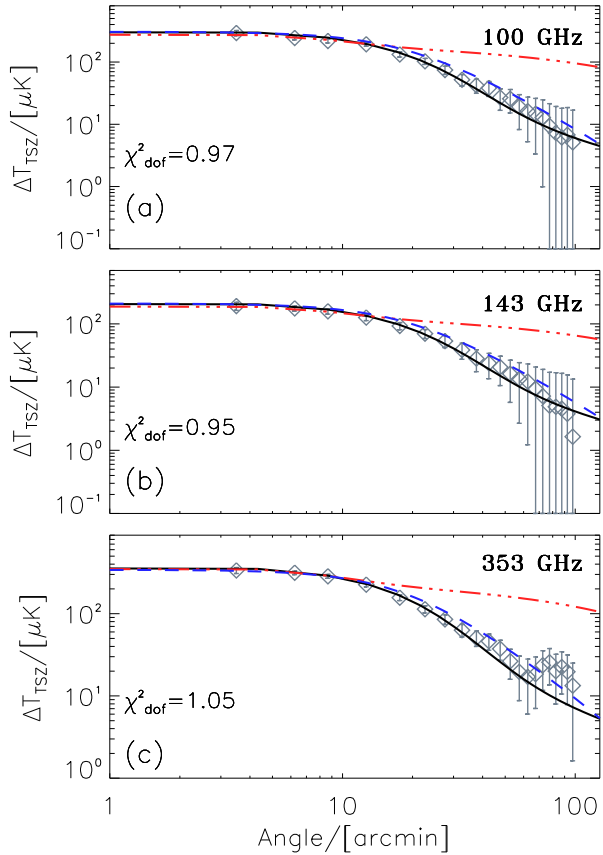}
\caption{Predicted and measured TSZ profile of the Coma cluster at different 
frequencies. For each channel, the MOG best fit model has been convolved 
with the antenna beam. The solid line represents the predicted model with the best fit values in Table \ref{tab:results},
while the red dot-dashed line shows the fitted profile with $\alpha$ and $\mu^{*^{-1}}$ fixed to their universal values.
Finally, the blue dashed line show the theoretical profile based on the Navarro-Frenk-White halo model with best
fit parameter from \cite{planck_int_X}.}\label{fig2}
\end{figure}

Once the MCMC algorithm has reach the convergence \cite{Gelman1992},  
we merged the four chains and computed the marginalized likelihood 
to constrain the model parameters. 
All results are summarized in Table~\ref{tab:results}, while in Fig. \ref{fig2}
we show the goodness of our fitting procedure.
\begin{table}
\begin{center}
\begin{tabular}{|l|c|}
\hline
 {\bf Parameter} & {\bf Results} \\
 \hline
 &\\[-0.2cm]
$P_c /[10^{-2}$ cm$^{-3}$ keV]  & $0.77\pm0.03$\\
 $\gamma$ & $1.40^{+0.15}_{-0.13}$\\  
 $\mu^{*^{-1}}$/[Mpc] & $4.22^{+0.55}_{-1.08}$\\
 $\alpha$ & $6.68^{+3.36}_{-2.08}$ \\
 \hline
\end{tabular}
\caption{Results from the MCMC.}\label{tab:results}
\end{center}
\end{table}

The Table summarizes same important results:
first, the parameter $\alpha$ is compatible at 68\% CL with its universal value $\alpha\simeq 8.89$
\cite{MoffatRahvar2013, MoffatRahvar2014}. Second,  the universal value of scale length $\mu^{*^{-1}}$ is
ruled out at more than $3.5\sigma$. Therefore, the assumption that the parameters of the Yukawa-potential can be 
assumed scale independent is also ruled out. Third, we find the polytropic index  
$\gamma=1.40^{+0.15}_{-0.13}$ to be consistent at $1.5\sigma$ level with 
the value  $\gamma\sim1.2$ preferred by observations and numerical simulations 
within the $\Lambda$CDM concordance model \cite{Ostriker2005, Ascasibar2006, Bode2009, capelo2012}.
Since the physical state of the gas in a galaxy cluster is determined by its
formation and evolution \cite{Bertschinger1985},
our results could be interpreted as an indication that MOG could be able to explain the emergence of the 
large scale structure, as well as the concordance model, if the theoretical parameter of the gravitational potential are free 
to vary. Finally, in Fig. \ref{fig2}, we plot the data (diamonds) with their associated error bars and 
the best fit  model (solid line). For comparison, we represent the fitted profile fixing $\alpha$ and $\mu^{*^{-1}}$ 
to their universal values (red dot-dashed line).  Panels (a-c) correspond to the three different  frequencies, while 
the $\chi^2$ per d.o.f, given in each panel, refers to the best fit model with all parameter free to vary. 
For the "universal" MOG profile we constrained: $P_c=(0.67\pm0.11)\times10^{-2}$ cm$^{-3}$ keV, and $\gamma=1.62^{+0.03}_{-0.49}$; 
it only fits the central region ($\lesssim 15$ arcminutes) of the galaxy cluster, while it  overestimates the TSZ emission
at larger apertures: at $\theta\sim 1$ degree the departure from the data is almost one order of magnitude. 
For comparison, we also show the SZ profile predicted using the generalized Navarro-Frenk-White profile (blue dashed line).
We used the best fit parameters specifically constrained for the Coma cluster by Planck Collaboration \cite{planck_int_X}.

\section{Further considerations on the universal nature of $(\alpha, \mu^*)$-parameters: 
limitation and future perspective of the analysis}\label{V}

Our results show a good agreement of $\alpha$ with its universal value but
a $3.5\sigma$ discrepancy in the scale length $\mu^{*^{-1}}$. 
The fact that $\mu^{*^{-1}}$ does not agree with its
universal fit could be interpreted as the consequence of the scale dependence of the modified potential:
$\Phi_{eff}(r \gg \mu^{*^{-1}})$ becomes Newtonian with an enhanced value of gravitational constant.
When assuming the universal MOG parameters  one fixes  $\mu^{*^{-1}}\sim$ kpc while
the typical scale length for a galaxy cluster is $\sim$ Mpc, thus
only $\alpha$ plays an important rule in to describe the gravitational interaction
and it fails to predict the TSZ profile at larger radii. 
Therefore, our results demonstrate that scale dependence of the MOG parameters 
play an important role at galaxy cluster scale and can not be neglected.

\begin{figure*}
\epsfxsize=1\textwidth \epsfbox{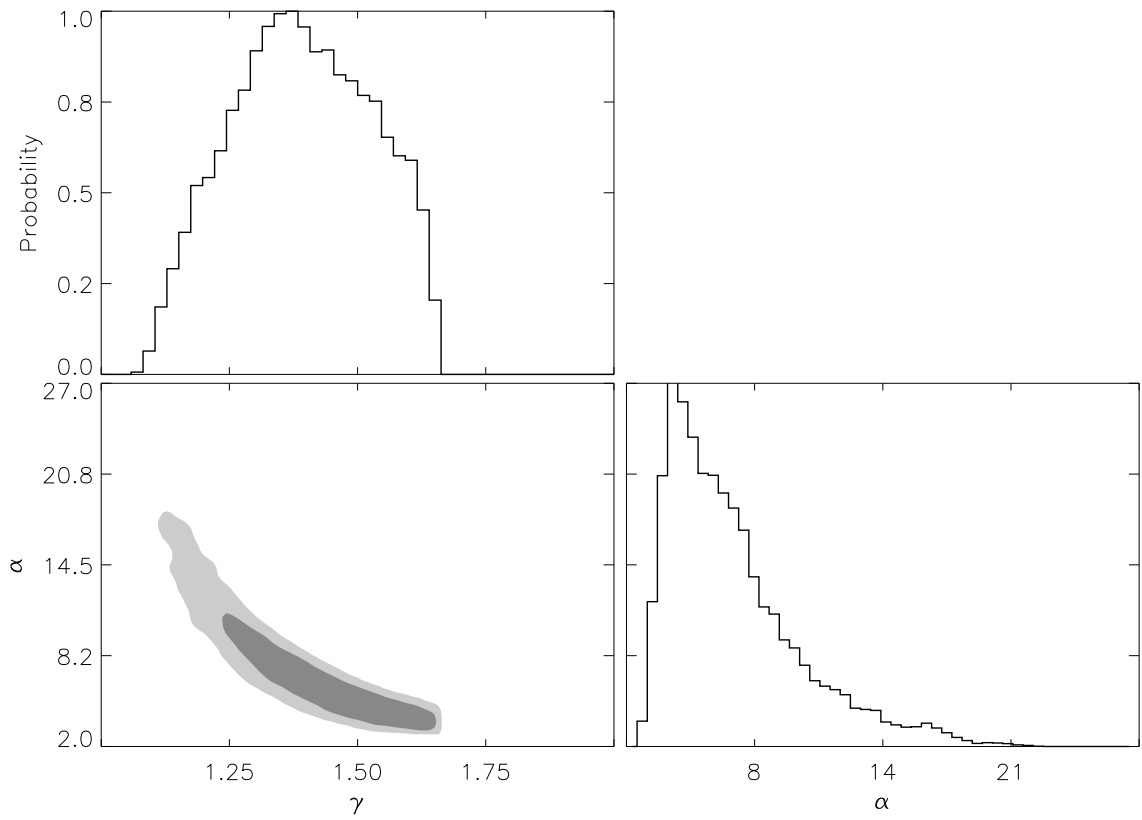}
\caption{2D marginalized contour of the pair of parameters $(\alpha, \gamma)$ obtained from the MCMC analysis.
For the pair of parameters the 68\% (dark gray) and 95\% CL (light gray),
the marginalized likelihood distributions are shown.}\label{fig3}
\end{figure*}

Another point of discussion is the assumptions of hydrostatic equilibrium and spherical symmetry
that are in our model. Although it has also been demonstrated that in the intermediate regions, 
where we are testing the model, both assumptions hold 
\cite{veritas2012, Terukina2014, Wilcox2015,Terukina2015},
it is well known that the presence of substructures, turbulences, heating and cooling processes
in the cluster core, and the departure from the spherical symmetry 
\cite{Coma_xray3, Coma_xray4, Coma_xray5, Coma_xray6, Coma_SB,  Coma_XT1, Coma_XT2, Gastaldello, FuscoFemiano} affect both 
the innermost and outermost regions of Coma cluster. The effect of such phenomena determines the physical state of the gas, 
and the degeneracy with the underlying theory of gravity. Actually, in our analysis we found a degeneracy between 
the gravitational parameter $\alpha$ and the polytropic index. This results is illustrated in Fig. \ref{fig3} 
where we plot the 2D marginalized contours obtained from our MCMC analysis.  
A way of studying the $\alpha-\gamma$ degeneracy  is including a non-thermal term in the pressure.
The proper strategy of doing this is to carry out hydrodynamical N-body simulations of each
specific MOG model, and compare the theoretical predictions to higher resolution data that
allow to resolve the cluster core  region ($<5$ arcminutes).  While we are currently limited by the angular 
resolutions of our foreground cleaned data (FWHM =10 arcminutes), the next-generation of full sky CMB missions such 
as COrE/PRISM \cite{PRISM} will have a much higher angular resolution ($\sim 3$ arcminutes) and 
frequency coverage (15 channels in the frequency range 45-795  GHz), 
and will allow to  properly investigate the relation between the underlying 
theory of gravity and the baryonic processes.

\section{Conclusions}\label{VI}

 We proposed an alternative test to probe the assumption of 
the universality of MOG weak field approximation \cite{MoffatRahvar2013, MoffatRahvar2014}.
 Despite the fact that, under this assumption, MOG theory is able to explain the phenomenology
at galactic scale, it is not clear if it is also able to describe the galaxy cluster. Therefore, there is an
important need to constrain the modified gravitational  potential in eq. \eqref{eq:MOG1} at galaxy cluster scales
in order to investigate if its scale dependence can be neglected or must be considered.
Thus, we  used the {\it Planck} 2013 Nominal maps to measure the TSZ temperature anisotropies 
on foreground cleaned patches centered at the position of Coma cluster. Then, we predict their
theoretical counterpart assuming that the gas was in hydrostatic equilibrium  
within the modified gravitational potential well (without Dark Matter), and it was well described 
by a polytropic equation of state. Finally,
we have employed a MCMC algorithm to fit our model to the measured TSZ profile, 
and summarized the best fit values of the model parameters in Table~\ref{tab:results}.
We found $\alpha$ to be consistent at the 68\% CL with its universal value \cite{MoffatRahvar2013},
while the scale length, $\mu^{*-1}$, was not compatible with such assumption at more than $3.5\sigma$. 
This latter result indicates a breakdown of the universality of the MOG weak field approximation demonstrating that,
in order to fit the TSZ temperature anisotropies of the Coma cluster, 
the scale dependence of the MOG parameters can not be neglected.

\section*{Acknowledgements}
We warmly acknowledge John Moffat for fruitful discussion and his valuable comments.
I.D.M acknowledge financial supports from University of the Basque Country UPV/EHU under the program
"Convocatoria de contrataci\'{o}n para la especializaci\'{o}n de personal 
investigador doctor en la UPV/EHU 2015", from the Spanish Ministerio de
Econom\'{\i}a y Competitividad through the research project FIS2010-15492,
and from  the Basque Government through the research project IT-956-16. 
M.\,D.\,L.\ is supported by the ERC Synergy Grant
``BlackHoleCam'' -- Imaging the Event Horizon of Black Holes (Grant
No.~610058). M.D.L. acknowledge INFN Sez. di Napoli (Iniziative Specifiche QGSKY and TEONGRAV).
This article is based upon work from COST Action CA1511 Cosmology and Astrophysics 
Network for Theoretical Advances and Training Actions (CANTATA), 
supported by COST (European Cooperation in Science and Technology).

\end{document}